\documentclass[preprintnumbers,prl,twocolumn]{revtex4}%
\usepackage{amssymb}
\usepackage{amsmath}
\usepackage{graphicx}%
\usepackage{amsfonts}
\renewcommand{\Im}{\mathop{\rm Im}\nolimits}

\begin{document}
\title{Thermally excited spin-current in metals with embedded ferromagnetic nanoclusters}
\author{Oleksandr Tsyplyatyev}
\affiliation{Physics Department, Lancaster University, Lancaster LA1 4YB, UK}
\author{Oleksiy Kashuba}
\affiliation{Physics Department, Lancaster University, Lancaster LA1 4YB, UK}
\author{Vladimir I. Fal'ko}
\affiliation{Physics Department, Lancaster University, Lancaster LA1 4YB, UK}

\begin{abstract}
We show that a thermally excited spin-current naturally appears 
in metals with embedded ferromagnetic nanoclusters. When 
such materials are subjected to a magnetic field, a spin
current can be generated by a temperature gradient across the sample 
as a signature of electron-hole symmetry breaking in a metal due to the electron
spin-flip scattering from polarised magnetic moments. Such a spin current can be observed via
a giant magneto-thermopower which tracks the polarisation state of the magnetic
subsystem and is proportional to the magnetoresistance. Our theory explains the recent experiment on Co clusters in copper by S. Serrano-Guisan \textit{et al} [Nature Materials AOP, doi:10.1038/nmat1713 (2006)]

\end{abstract}
\maketitle

Theoretical studies of spin-currents in metals and semiconductors have recently
been fuelled by the search for materials to develop spintronic devices \cite{Spintronics}.
 Among the  materials studied of  particular interest
had been composite ferromagnetic and nonmagnetic  metallic systems, such as
metallic multilayers \cite{FertMR} and metals with embedded ferromagnetic
nanoclusters \cite{ZhangLevy}, where the electronic transport strongly depends on
the electron spin leading to the effect of the giant magnetoresistance (MR).

In metals with ferromagnetic nanoclusters \cite{ZhangLevy,MR} or magnetic
impurities~\cite{Yosida} with a large spin, $s\gg1$, the magnetoresistance
reflects the degree of polarisation, $\langle l_{z}\rangle=\coth\left(
y\right)  -y^{-1}$, $y=\mu Bs/kT$ of a magnetic sub-system by a magnetic field
$\mathbf{B}=\mathbf{n}_{z}B$,%
\[
\Delta(B)\equiv\frac{R(B)}{R(0)}-1=-\frac{\left(  \tau_{{\small \uparrow}%
}-\tau_{{\small \downarrow}}\right)  ^{2}}{4\tau^{2}}\langle l_{z}\rangle^{2}.
\]
The MR effect develops across the magnetic field range $B_{MR}\sim
kT/\mu s$, and its strength depends on the relative difference
between the mean free path time $\tau_{{\small
\uparrow(\downarrow)}}$ of electrons with spins
parallel(antiparallel) to the cluster polarisation axis. Here,
$\tau_{{\small \uparrow}\left(\small \downarrow\right)}$ is
defined as the mean free path in a metal with magnetic clusters
whose spin polarisations are fully up(down). The Fermi density of
states per spin, $\gamma$, is identical for both electron spin
orientations, and $\tau=\frac{1}{2}\left( \tau _{{\small
\uparrow}}+\tau_{{\small \downarrow}}\right)$ is a characteristic
mean free path time, while $\mu$ is the Bohr magneton.

In this Letter we show that a thermally excited spin-current
naturally appears in metals with embedded ferromagnetic
nanoclusters (FmnC's) when subjected to a magnetic field and
temperature gradient. This effect appears through the
manifestation of electron-hole symmetry breaking via electron
spin-flip scattering from polarised magnetic moments. When two
parts of a metallic sample with embedded FmnC's are held at
different temperatures $T_{1}>T_{2}$,\ the thermally equilibrating
heat flux must be accompanied by a transport of magnetisation
\cite{Korenblit,Fert,McCann}, \textit{i.e.} a spin current,
$\mathbf{j}_{s}$. This spin current acts to equilibrate the
cluster polarisation on opposing sides of the temperature
gradient. Locally, the equilibration process requires spin
transfer from clusters to conduction electrons via spin-flip
scattering at a rate $\tau_{s}^{-1}$, which is the
inverse mean free path time of the spin-flip scattering process (formally defined latter, Eq.(\ref{taus})). After a
single spin-flip scattering event, the scattered electron will
carry the transferred magnetisation while diffusing (with
diffusion coefficient $D$) between different parts of the sample
held at different temperatures, thus leading to a spin-current,
\begin{equation}
\mathbf{j}_{s}=\beta\mathbf{\nabla}T,\;\;\beta\approx\tfrac{1}{2}\hbar k\gamma
D\,\frac{\tau f}{\tau_{s}}\,\langle l_{z}\rangle.\label{spin-current}%
\end{equation}
Here $f=\frac{x^{2}e^{x}}{(e^{x}-1)^{2}}$ and $x=\mu B/kT$ show
that the spin current will persist up to a value of the magnetic
field $B_s\sim kT/\mu$, which is much larger than the typical
field value at which MR develops, $B_{MR} \ll B_s$. Additionally,
due to the difference in the mean free path times of ``up'' and
``down'' spin carriers, $\left( \tau_{{\small \uparrow}}-\tau
_{{\small \downarrow}}\right) \,\langle l_{z}\rangle$, the spin
current `drags' a charge current,
$\mathbf{j}\propto\varkappa\mathbf{\nabla}T$. In an open circuit,
this generates a thermopower, $V_{12}\propto
c\left(B\right)[T_{1}-T_{2}]$, with a strong magnetic field
dependence resembling (for $B<kT/\mu$) that of MR,
\begin{equation}
\Xi(B)\equiv
c(B)-c(0)\approx\frac{k}{e}\,\frac{\tau^{2}f}{\tau_{s}\left(
\tau_{{\small \downarrow}}-\tau_{{\small \uparrow}}\right)
}\,\Delta
(B).\label{eq:thermp_approx}%
\end{equation}

A microscopic justification of the above-presented
phenomenological argument is supported by the following analysis
of the electron-hole asymmetry in a composite metal with
ferromagnetic components. \ In normal metals, the electron-hole
asymmetry leading to the thermopower is caused by the energy
dependence of the density of states of electrons near the Fermi
level. In materials containing a polarised magnetic subsystem, the
electron-hole asymmetry is created in an alternative way --- via
the energy and spin dependence of a quasi-particle scattering rate
\cite{Korenblit,Fert,McCann}. In the system we discuss in this
Letter, the formation of kinetic electron-hole asymmetry can be
illustrated using  Kondo-type model \cite{Kondo} which treats a
FmnC as an impurity with a large spin, $s$, and contains all the
necessary ingredients to describe the MR effect simultaneously
with the generation of spin-current and magneto-thermopower.

Thus, we model the FmnC's by the Hamiltonian
\[
-\mu B\hat{s}_{z}+\int\psi^{\dagger}(\mathbf{r})\sum_{i}(U+J\mathbf{\hat
{\sigma}\cdot\hat{s}})\delta(\mathbf{r}-\mathbf{r}_{i})\psi(\mathbf{r}%
)d^{3}\mathbf{r}.
\]
Here, the potential $U$ accounts for the FnmC charge and for band
mismatch between normal and magnetic metals, whereas $J$ is the
exchange interaction; $\mu s$ is the magnetic moment of each
cluster; $\mathbf{\sigma}$ stands for the electron spin operator.
The MR analysis in materials with large spin clusters, $s\gg1$ can
be done using a static exchange field model \cite{ZhangLevy},
where the operator $\mathbf{\hat{s}}$ is replaced by $s\mathbf{l}$
(here,$\mathbf{\ l}$ is a unit vector in the direction of
polarisation of an individual cluster). In contrast, the analysis
of a thermally excited spin current and the magneto-thermopower
needs to take into account the quantum nature of the cluster spin,
manifested through the electron spin-flip process.

\begin{figure}
[t]
\begin{center}
\includegraphics[
height=2.2701in,
width=3.1021in
]%
{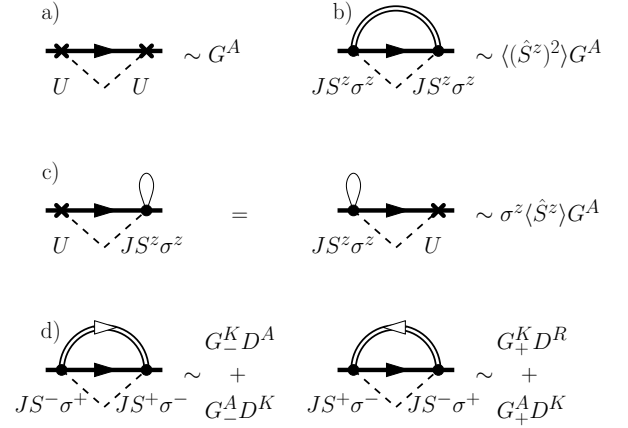}%
\caption{{\small Self-energy digrams: thick solid line is free electrons
Green's function, }$G_{\pm}=G(\varepsilon\pm\mu B)${\small , thin dashed line
is impurity line, thin solid line connected both ends to a single vertex is an
average of spin operator, double thin solid line connected to two vertices is
the spin-spin correlator, cross is the potential vertex $U$ and dot is the
exchange vertex $J$.}}%
\label{fig:diags}%
\end{center}
\end{figure}

The formation of electron-hole asymmetry in the quasi-particle
lifetime becomes apparent after analyzing the imaginary part of
the self-energy to the lowest non-vanishing order of perturbation
theory. Fig. \ref{fig:diags} shows the Keldysh diagrams which
appear in the self-energy $\hat{\Sigma}^{A(R)}$ in the second
order of the electron-cluster interaction, after averaging over
random positions, $\mathbf{r}_{i}$, of the clusters. We assume a
thermal distribution of cluster spin polarisation, with
temperature $T$. The diagrams in Fig.~\ref{fig:diags} incorporate
two types of vertices: crosses stand for the potential $U$, and
dots for the exchange
coupling $J$. Dashed lines represent impurity averaging [$\mathbf{--}%
=n_{c}\delta(\mathbf{r}-\mathbf{r}^{\prime})$]. Thin solid lines
indicate the thermodynamic average and dynamical correlators of
the cluster spin: a loop stands for the thermodynamic average
$\langle s_{z}\rangle$, a double line without an arrow is
$\langle(s_{z})^{2}\rangle$, whereas a thin double line with
arrows stands for the free-spin Green functions \cite{giovannini},
$D^{R/A}(t)=\mp
i\langle\lbrack\hat{s}^{+}(t)\hat{s}^{-}(0)]_{-}\rangle
\theta(\pm t)$ and $D^{K}(t)=-i\langle\left\{  \hat{s}^{+}(t)\hat{s}%
^{-}(0)\right\}  _{+}\rangle$, where $\hat{s}^{\pm}=\hat{s}_{x}\pm i\hat
{s}_{y}$. In the energy representation,
\begin{multline}
\begin{picture}(22,8) \put(0,3){\line(1,0){7}} \put(0,5){\line(1,0){7}} \put(4,0){ \begin{picture}(10,8) \put(0,0.5){\line(0,1){6.5}} \put(10,3.75){\line(-3,-1){10}} \put(10,3.75){\line(-3,1){10}} \end{picture}} \put(15,3){\line(1,0){7}} \put(15,5){\line(1,0){7}} \end{picture}\,=D^{R/A}%
(\omega)=\frac{2\langle s_{z}\rangle}{\omega-\mu B\pm i\delta},\\
D^{K}(\omega)=\coth\frac{\omega}{2T}\left(  D^{R}(\omega)-D^{A}(\omega
)\right)  .
\end{multline}
Although Wick's theorem cannot be applied to spin operators in higher orders
of perturbation theory, this does not affect the analysis of the diagrams in
the Born approximation. Finally, bold solid lines with arrows are free electronic
Green functions,
\begin{multline}
\begin{picture}(22,8) \put(7,0){\begin{picture}(10,8) \linethickness{1.5pt} \put(0.75,0.75){\line(0,1){7}} \multiput(10.5,4)(0,0.1){15}{\line(-3,-1){10}} \multiput(10.5,4.5)(0,-0.1){15}{\line(-3,1){10}} \end{picture}} \linethickness{2pt} \multiput(0,4)(0,1){2}{\line(1,0){20}} \end{picture}\,=G^{R/A}%
(\varepsilon,\mathbf{p})=\frac{1}{\varepsilon-\xi_{\mathbf{p}}\pm i\delta},\\
G^{K}(\varepsilon,\mathbf{p})=\tanh\frac{\varepsilon}{2T}\left(
G^{R}(\varepsilon,\mathbf{p})-G^{A}(\varepsilon,\mathbf{p})\right)  ,
\end{multline}
where the electron energy $\varepsilon$ is determined with respect to the
Fermi level.%

The diagram in Fig.~\ref{fig:diags} (a) describes scattering from
a scalar potential, whereas diagrams (b) and (d) only include the
exchange interaction. The interference between the scalar and
exchange potential scattering amplitudes (from the same impurity)
is taken into account by the diagrams in Fig. 1(c), which is
responsible for the spin-dependence of the electron mean free path
in the presence of polarised clusters. Together, these diagrams yield \cite{Kondo,Glazman}%
\begin{multline}
\Im\hat{\Sigma}^{A}(\varepsilon)=\pi\gamma n_{c}\left[  U^{2}+J^{2}%
s(s+1)+2UJ\langle s_{z}\rangle\hat{\sigma}_{z}\right]  -\\
-\pi\gamma n_{c}J^{2}\langle s_{z}\rangle\hat{\sigma}_{z}{\tanh\left(
\frac{\varepsilon+\mu B\hat{\sigma}_{z}}{2T}\right)  },\label{eq:imse}%
\end{multline}
where $n_{c}$ is the concentration of FmnC's. Diagrams (a)--(c)
describe elastic processes. The spin-flip diagrams (d) contain an
inelastic part resulting in the energy-dependent contribution
towards $\Im\hat{\Sigma}^{A}$ in Eq.(\ref{eq:imse}), and they are responsible for the kinetic
electron-hole asymmetry. This asymmetry is most pronounced when
the ensemble of clusters is fully polarised, $\left\langle
l_{z}\right\rangle \rightarrow1$, and vanishes when $\left\langle
l_{z}\right\rangle =0$.

\begin{figure}
[ptb]
\begin{center}
\includegraphics[
height=3.0364in,
width=2.6091in
]%
{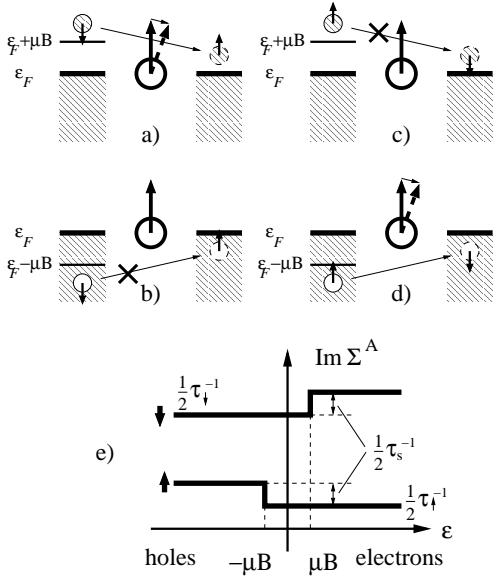}%
\caption{{\small Schematic representation of allowed and forbidden processes
leading to the energy and spin dependent quasiparticles relaxation rate ( a),
d) - allowed processes, b), c) - forbidden processes, e) - dependence of
$\Im\Sigma^{A}$ as function of energy for different spin directions - graph is
drawn for the case of $JU<0$)}}%
\label{fig:scattering}%
\end{center}
\end{figure}

To explain the origin of this asymmetry and its polarisation
dependence we consider the limit of $\left\langle
l_{z}\right\rangle \rightarrow1$, so that all FmnC's are polarised
`up'. Fig. 2(a) shows that an incident spin-${\small \downarrow}$
electron with energy $\epsilon>\mu B$ above the Fermi level is
able to flip its spin turning the spin of FmnC from the
polarisation axis. The amplitude of such a process is $A\sim
J\langle s-1|\hat{s}^{-}|s\rangle\sim J\sqrt{s}$, which results in
the scattering rate $\tau_{s}^{-1}\propto J^{2}s$. The relevant
range of a quasiparticle excitation energy for this process is set
by the FmnC energy splitting, $\mu B$, between its initial and
scattered spin state, which differ by $s_{z}-s_{z}^{\prime}=1$. A
similar process is possible for the incident spin-${\small
\uparrow}$ hole with $\epsilon<-\mu B$ below the Fermi level, Fig.
2(d): it corresponds to the reverse process of an equilibrium
spin-${\small \downarrow}$ electron relaxing into an empty state
with opposite spin below the Fermi level (hole). \ In
contrast, neither spin-${\small \uparrow}$ electrons above the
Fermi level, nor spin-${\small \downarrow}$ holes shown in Figs.
2(b,c) can scatter with changing spin state, since they cannot
increase further the maximal $s_{z}=s$ of a cluster. All together,
these processes determine the energy-dependent part in the
electron scattering rate from fully polarized clusters,
\begin{align}
\Im\hat{\Sigma}_{\left\langle l_{z}\right\rangle \rightarrow1}^{A}  & =%
\begin{pmatrix}
\frac{\tau_{\uparrow}^{-1}+\theta(-\varepsilon-\mu B)\tau_{s}^{-1}}{2} & 0\\
0 & \frac{\tau_{\downarrow}^{-1}+\theta(\varepsilon-\mu B)\tau_{s}^{-1}}{2}%
\end{pmatrix}
,\nonumber\\
\tau_{{\small \uparrow(\downarrow)}}^{-1}  & =\frac{2\pi n_{c}\gamma}{\hbar
}(U^{2}+J^{2}s(s+1)\pm2UJs),\ \nonumber\\
\tau_{s}^{-1}  & =\frac{4\pi n_{c}\gamma}{\hbar}J^{2}s.\label{taus}%
\end{align}
Note that for a cluster with $N$ magnetic atoms $S\propto N$ and $U\propto N$ so that
 $\tau^{-1}_{\uparrow(\downarrow)}\propto N^2$, whereas $\tau^{-1}_s \propto N$. 
The resulting behaviour of the quasi-particle lifetime for electrons in the
presence of polarised FmnC's is sketched in Fig. \ref{fig:scattering}(e). It
indicates that the electron-hole asymmetry is inverted in the opposite spin
channels, in accordance with the symmetry of the Kondo problem
\cite{KondoBook}.

The inverted electron-hole asymmetry in the opposite spin channel
results in the generation of a spin current when two parts of the
system are held at different temperatures. To describe a thermally
generated spin-current, $\mathbf{j}_{s}=\beta\mathbf{\nabla}T$, as
well as the electric current $\mathbf{j}=\varkappa\nabla T$, we
use the kinetic equation approach. That is, we study a
steady-state kinetic equation, \
\begin{equation}
\left[
\mathbf{v}\cdot\nabla-e\mathbf{E}\cdot\partial_{\mathbf{p}}\right]
\varrho
_{\pm}=\left\langle I_{\pm}(\hat{\varrho},s_{z})\right\rangle ,\label{kineq}%
\end{equation}
where the collision term $I_{\pm}(\hat{\varrho},s_{z})$ describes
the balance in the distribution functions $\varrho_{+}$ of
spin-${\small \uparrow}$ and $\varrho_{-}$ of spin-${\small
\downarrow}$ electrons which scatter from a group of FmnC's in a
given initial spin state, $s_{z}$. The brackets
$\left\langle\ldots\right\rangle$ stand for averaging over the
thermal distribution of cluster spins. The averaged collision term
therefore takes into account all electron scattering processes
described by the diagrams in
Fig. \ref{fig:diags},%
\begin{multline*}
I_{\pm}=\frac{2\pi n_{c}\left\vert U\pm Js_{z}\right\vert ^{2}}{\hbar}%
\int\frac{d^{3}\mathbf{p}^{\prime}}{h^{3}}\left(  \varrho_{\pm}^{\prime
}-\varrho_{\pm}\right)  \delta\left(  \epsilon_{\mathbf{p}}-\epsilon
_{\mathbf{p}^{\prime}}\right) \\
+\frac{2\pi n_{c}J^{2}}{\hbar}\int\frac{d^{3}\mathbf{p}^{\prime}}{h^{3}%
}\left[  \left(  s^{2}-s_{z}^{2}+s\pm s_{z}\right)  \varrho_{\mp}^{\prime
}\left(  1-\varrho_{\pm}\right)  \right. \\
\left.  -\left(  s^{2}-s_{z}^{2}+s\mp s_{z}\right)  \varrho_{\pm}\left(
1-\varrho_{\mp}^{\prime}\right)  \right]  \delta\left(  \epsilon
_{\mathbf{p}^{\prime}}-\epsilon_{\mathbf{p}}\mp\mu B\right)  .
\end{multline*}
The first term in $I_{\pm}$ describes the elastic spin-conserving processes in
Fig. \ref{fig:diags}(a-c), whereas the second takes care of spin-flip
processes corresponding to diagrams in Fig.~\ref{fig:diags}(d). By inspection,
one can see that the probabilities of electrons ${\small \downarrow
}\rightarrow{\small \uparrow}$ and ${\small \uparrow}\rightarrow
{\small \downarrow}$ spin-flip scattering from the same cluster with a given
$s_{z}$ slightly differ. In a system where FmnC's are partially polarised
(along the external magnetic field $\mathbf{B}=\mathbf{n}_{z}B$) this leads to
the energy- and spin-dependent electron momentum relaxation rate, $2\Im
\Sigma_{\alpha\alpha}^{A}/\hbar$ (where $\alpha={\small \uparrow,\downarrow}$)
given by Eq. (\ref{eq:imse}). \ In the leading order in a small temperature
gradient ($v_{F}\tau T^{-1}\nabla T\ll1$), the spin-current, $\mathbf{j}%
_{s}\mathbf{=}\frac{\hbar}{6}\int\gamma d\varepsilon\mathrm{Tr}\{\hat{\sigma
}_{z}\hat{\boldsymbol{\rho}}\}$ and electric current, $\mathbf{j=}\frac{e}%
{3}\int\gamma d\varepsilon\mathrm{Tr}\{\hat{\boldsymbol{\rho}}\}$ are
determined by the first angular harmonic in momentum space, $\hat
{\boldsymbol{\rho}}$ of the electron spin-density matrix, $\hat{\varrho
}\approx\varrho_{T}+\hat{\boldsymbol{\rho}}\cdot\mathbf{p}/p$. This first
angular harmonic can be found using the equation
\[
\frac{2}{\hbar}\Im\hat{\Sigma}^{A}\boldsymbol{\hat{\rho}}=\frac{\varepsilon
v_{F}\partial_{\varepsilon}\varrho_{T}}{T}\nabla T\boldsymbol{\,},
\]
where $\varrho_{T}(\varepsilon)=\left[  e^{\varepsilon/kT}+1\right]  ^{-1}$ is
the Fermi function.

As a result, we calculate the spin-current,\ $\mathbf{j}_{s}=\beta
\mathbf{\nabla}T$, which, for large spin clusters, $s\gg1$ is described by
\begin{align}
\beta &  =-\frac{2\hbar k\gamma v_{F}^{2}}{3}\frac{\tau_{{\small \uparrow}%
}^{2}\tau_{{\small \downarrow}}^{2}\,f}{\tau_{s}}\,\langle l_{z}%
\rangle\,\times\label{spin-current1}\\
&  \;\;\;\;\;\;\;\;\;\times\,\frac{\left(  \tau_{{\small \uparrow}}%
+\tau_{{\small \downarrow}}\right)  ^{2}+\left(  \tau_{{\small \uparrow}}%
-\tau_{{\small \downarrow}}\right)  ^{2}\langle l_{z}\rangle^{2}}{\left[
\left(  \tau_{{\small \uparrow}}+\tau_{{\small \downarrow}}\right)
^{2}-\left(  \tau_{{\small \uparrow}}-\tau_{{\small \downarrow}}\right)
^{2}\langle l_{z}\rangle^{2}\right]  ^{2}}\,.\nonumber
\end{align}
In the above result, scattering rates $\tau_{{\small \uparrow(\downarrow)}%
}^{-1}$ and $\tau_{{\small s}}^{-1}$ are defined in Eq. (\ref{taus}), and the
factor $f=\frac{x^{2}e^{x}}{(e^{x}-1)^{2}}$ with $x=\mu B/kT$ is specified for
the case $s\gg1$ and takes into account the suppression of the spin transfer
rate at low temperatures and high magnetic field, such that $\mu B>kT$, since
it requires the electron energy transfer $\epsilon-\epsilon^{\prime}=\mu B$.
Also, in most of the metals with embedded ferromagnetic clusters, the maximum
MR effect is $\Delta\lesssim10\%$, which is additionally suppressed by
non-magnetic impurities and phonon scattering. Therefore, we can simplify the
result in Eq. (\ref{spin-current1}) further, using the fact that $\left(
\tau_{{\small \uparrow}}-\tau_{{\small \downarrow}}\right)  /\left(
\tau_{{\small \uparrow}}+\tau_{{\small \downarrow}}\right)  \ll1$, which leads
to the approximate form in Eq. (\ref{spin-current}).

Due to the difference, $\left(  \tau_{{\small \uparrow}}-\tau
_{{\small \downarrow}}\right)  \,\langle l_{z}\rangle$ between the
mean free path times of spin-${\small \uparrow}$ and -${\small
\downarrow}$ carriers scattering from the ensemble of partially
polarised FmnC's, the spin-current `drags' a charge current
$\mathbf{j}\propto\varkappa\mathbf{\nabla}T$. For $s\gg1$, we find
\[
\varkappa=\frac{8ev_{F}^{2}\gamma k}{3\tau_{s}}\,\frac{\tau_{{\small \uparrow
}}^{2}\tau_{{\small \downarrow}}^{2}\left(  \tau_{\uparrow}-\tau_{\downarrow
}\right)  \,\left\langle l_{z}\right\rangle ^{2}\,f\left(  \frac{\mu B}%
{kT}\right)  }{\left[  \left(  \tau_{\uparrow}+\tau_{\downarrow}\right)
^{2}-\left(  \tau_{\uparrow}-\tau_{\downarrow}\right)  ^{2}\left\langle
l_{z}\right\rangle ^{2}\right]  ^{2}}\,.
\]
Together with the MR of this material obtained using the same approximations,
\[
R=\frac{3}{e^{2}v_{F}^{2}\gamma}\,\frac{\tau_{\uparrow}+\tau_{\downarrow}%
}{4\tau_{{\small \uparrow}}\tau_{{\small \downarrow}}}\,\left[  1-\frac
{\left(  \tau_{\uparrow}-\tau_{\downarrow}\right)  ^{2}}{\left(
\tau_{\uparrow}+\tau_{\downarrow}\right)  ^{2}}\left\langle l_{z}\right\rangle
^{2}\right]  ,
\]
this determines the thermopower coefficient $c=-\varkappa R$ with
a magnetic field dependence. The magneto-thermopower $\Xi(B)\equiv
c(B)-c(0)$ is related to the observable MR as
\begin{equation}
\Xi(B)=\frac{k}{e}\frac{\tau_{{\small \downarrow}}\tau_{{\small \uparrow}}%
}{\tau_{s}\left(  \tau_{{\small \downarrow}}-\tau_{{\small \uparrow}}\right)
}\frac{\Delta(B)}{1+\Delta(B)}f\left(  \frac{\mu B}{kT}\right)
.\label{eq:answer}%
\end{equation}
The magnetic field dependence of $\Xi(B)$, implicit in the above
result, contains two field scales \cite{spin-1/2}. \ At a
low magnetic field where the polarisation of clusters develops,
the MTP is proportional to the MR and saturates together with
$\Delta(B)$ at the field $B_{\mathrm{MR}}$. At a higher
field range, $B>kT/\mu$ the MTP is suppressed and dies away. For
weak MR materials, $\Delta\lesssim0.1$, where $\tau_{{\small \downarrow}}%
-\tau_{{\small \uparrow}}\ll\tau=\frac{1}{2}\left(  \tau_{{\small \downarrow}%
}+\tau_{{\small \uparrow}}\right)$, the above relation can be
approximated using the formula in Eq. (\ref{eq:thermp_approx}).

At the early stage of study of Kondo impurity in metals\cite{KondoBook} a strong 
magneto-thermopower has been noticed in various
dilute magnetic alloys \cite{Huntley,Kopp,Guenault,Azevedo}. Later, a strong magnetic field
dependent thermopower has been observed in metals with embedded
ferromagnetic nanoclusters \cite{Salamon,Piraux,Sato,Kobayashi,Sakurai},
where it has been attributed to the energy dependent density of
states in a weakly ferromagnetic metal with a complex band
structure \cite{Interpret}. However, when viewed in terms of the above-presented theory,
these observations indicate the presence of thermally excited
spin-current, Eqs. (\ref{spin-current},\ref{spin-current1})
generated by a temperature gradient. 

Recently, a giant MTP has been measured 
in copper with embedded Co clusters of a controlled size. It was noticed that MTP weakens
with increase of the number of atoms, $N$, in a FmnC\cite{Serrano_Guisan}. This behaviour as well as the observed direct correspondence between MTP and MR can be explained on 
the basis of the result in Eq.(\ref{eq:answer}). The matter is that for a cluster 
with $N$ magnetic atoms $\tau^{-1}_{\uparrow(\downarrow)}\propto N^2$, whereas 
$\tau^{-1}_s\propto N$. Therefore, for two samples with the same densities of 
the embedded clusters containing $N$ and $N'$ atoms, we estimate
$\Xi(N)/\Xi(N')\propto N'/N$, the MTP is weaker in the system with larger clusters.

The authors thank B. Altshuler, G. Bauer, C. Marrows, L.Glazman,
A. Guenault, E. McCann and J. Jefferson for
discussions. This work was funded by the EU STREP
NPM2--CT-2003-505587 'SFINX' and the Lancaster-EPSRC Portfolio
Partnership EP/C511743.


\begin{thebibliography}{99}                                                                                               %
\bibitem {Spintronics}I. \v{Z}uti\'{c}, J. Fabian, and S. Das Sarma, Rev. Mod.
Phys. \textbf{76}, 323-410 (2004); G.A. Prinz, Science \textbf{282}, 1660
(1998) and refs. therein

\bibitem {FertMR}T. Valet and A. Fert, Phys. Rev. B \textbf{48}, 7099 (1993);
D.H. Mosca \textit{et al}, Journ. of Magn. and Magn. Mat. \textbf{94}, L1 (1991)

\bibitem {ZhangLevy}S. Zhang, Appl. Phys. Lett. \textbf{61}, 1855 (1992); S.
Zhang and P.M. Levy, J. Appl. Phys. \textbf{73}, 5315 (1993); H. Camlong, S.
Zhang, and P.M. Levy, J. Appl. Phys. \textbf{75}, 6906 (1994), P. Holody et
al, Phys. Rev. B \textbf{50}, 12999 (1994)

\bibitem {MR}A. Berkowitz \textit{et al.}, Phys. Rev. Lett. \textbf{68}, 3745
(1992); J. Xiao \textit{et al.}, Phys. Rev. B \textbf{46}, 9266 (1992)

\bibitem {Yosida}K. Yosida, Phys. Rev. \textbf{107}, 396 (1957); H. Rohrer,
Phys. Rev. \textbf{174}, 583 (1968)

\bibitem {Korenblit}I. Korenblit, J. Phys. F \textbf{12}, 1259 (1982)

\bibitem {Fert}L. Piraux \textit{et al}, Journ. of Magn. and Magn. Mat.
\textbf{110}, L247 (1992); J.L. Duval \textit{et al}, J. Appl. Phys.
\textbf{75}, 7070 (1994)

\bibitem {McCann}E. McCann and V.I. Fal'ko, Appl. Phys. Lett. \textbf{81},
3609 (2002); E. McCann and V.I. Fal'ko, Phys. Rev. B \textbf{66}, 134424 (2002)

\bibitem {Interpret}L. Xing \textit{et al}, Phys. Rev. B \textbf{48}, 6728 (1993)

\bibitem {giovannini}B. Giovannini and S. Koide, Progr. Theor. Phys.
\textbf{34}, 705 (1965)

\bibitem {Kondo}J. Kondo, Progr. Teor. Phys. \textbf{34}, 372 (1965);
L.Gurevich and I. Yassievich, Sov. Phys. JETP \textbf{20}, 922 (1965); K.Maki,
Progr. Teor. Phys. \textbf{41}, 586 (1969); R.Weiner and M.Beal-Monod, Phys.
Rev. B \textbf{2}, 2675 (1970); N. Kawakami, T. Usuki, and A. Okiji, J. Phys.
Soc Jpn 56, 1539 (1987)

\bibitem {Glazman}M.G. Vavilov, L.I. Glazman, and A.I. Larkin, Phys. Rev. B
\textbf{68}, 075119 (2003)

\bibitem {majorana}M. Peskin, D. Schroeder,\textit{\ Introduction to Quantum
Field Theory}, HarperCollins 1995

\bibitem {KondoBook}A. Hewson, \textit{The Kondo Problem to Heavy Fermions},
Cambridge University Press, 1993

\bibitem {spin-1/2}For a metal with spin-$\frac{1}{2}$ impurities, their
polarisation and MR develop over the same range of a magnetic field as the
field suppressing the electron spin-flip scattering. Therefore, in a metal
with spin-$\frac{1}{2}$ Anderson impurites, magneto-thermopower would display
a sharp non-monotonic field dependence with a maximum effect at $B\approx
2kT/\mu$, $\Xi=\frac{k}{e}\frac{\tau_{{\small \downarrow}}\tau
_{{\small \uparrow}}}{\tau_{s}\left(  \tau_{{\small \downarrow}}%
-\tau_{{\small \uparrow}}\right)  }\frac{\Delta(B)}{\cosh^{2}\left(  \mu
B/2kT\right)  }.$

\bibitem {Huntley}D. Huntley and C. Walker, Can. J. Phys. \textbf{47}, 805 (1969)

\bibitem {Kopp}R. Berman \textit{et al}, Physics Letters A \textbf{27}, 464
(1968); R. Berman and J. Kopp, J. Phys. F \textbf{1}, 457(1971); J. Kopp,
\textit{ibid.} 6, 1211 (1975)

\bibitem {Guenault}M. Read and A. Guenault, J. Phys. F \textbf{4}, 94 (1974)

\bibitem {Azevedo}L. Azevedo \textit{et al}, Phys. Rev. B \textbf{20}, 4450 (1979)

\bibitem {Salamon}J. Shi, E. Kita, L. Xing, and M. Salamon, Phys. Rev. B
\textbf{48}, 16119 (1993); J. Shi \textit{et al}, Phys. Rev. B \textbf{54},
15273 (1996)

\bibitem {Piraux}L. Piraux \textit{et al}, Phys. Rev. B \textbf{48}, 638 (1993)

\bibitem {Sato}H. Sato \textit{et al}, Journ. of Magn. and Magn. Mat.
\textbf{152}, 109 (1996); H.Sato, Mat. Sci. Eng. B \textbf{31}, 101 (1995);
H.Sato \textit{et al}, J. Phys. Condens. Matter \textbf{7}, 7053 (1995)

\bibitem {Kobayashi}Y. Kobayashi \textit{et al}, J. Phys. Cond. Matter
\textbf{8}, 11105 (1996)

\bibitem {Sakurai}J. Sakurai \textit{et al}, J. Phys. Soc. Jpn. \textbf{66},
2240 (1997)
\bibitem{Serrano_Guisan} S. Serrano-Guisan \textit{et al}, Nature Materials AOP, doi:10.1038/nmat1713 (2006)

\end{thebibliography}
\end{document}